# Direct dry transfer of chemical vapor deposition graphene to polymeric substrates

*Guilhermino J. M. Fechine *[#1], Iñigo Martin-Fernandez [#2,3], George Yiapanis[4], Ricardo V. Bof de Oliveira[5], Xiao Hu[6], Irene Yarovsky[4], Antônio H. Castro Neto[2,3,7], Barbaros Özyilmaz*[2,3,7]*

[1] Graphene and Nano-materials Research Center – Mackgraphe, Mackenzie Presbyterian University, São Paulo, Brazil.

[2] Graphene Research Centre, National University of Singapore, 6 Science Drive 2, Singapore 117546.

[3] Department of Physics, National University of Singapore, 2 Science Drive 3, Singapore 117542.

[4] School of Aerospace, Mechanical and Manufacturing Engineering, RMIT University, Melbourne, Australia

[5] Institute of Chemistry, Universidade Federal do Rio Grande do Sul, Porto Alegre, Brazil

[6] School of Materials Science and Engineering, Nanyang Technological University - NTU, Singapore

[7] Graduate School for Integrative Sciences and Engineering (NGS), National University of Singapore, 28 Medical Drive, Singapore 117456.

[#]These authors contributed equally to this work.

*Abstract*

We demonstrate the direct dry transfer of large area Chemical Vapor Deposition graphene to several polymers (low density polyethylene, high density polyethylene, polystyrene, polylactide acid and poly(vinylidenefluoride-co-trifluoroethylene)) by means of only moderate heat and pressure, and the later mechanical peeling of the original graphene substrate. Simulations of the graphene-polymer interactions, rheological tests and graphene transfer at various experimental conditions show that controlling the graphene-polymer interface is the key to controlling graphene transfer. Raman spectroscopy and Optical Microscopy were used to identify and quantify graphene transferred to the polymer substrates. The results showed that the amount of graphene transferred to the polymer, from no-graphene to full graphene transfers, can be achieved by fine tuning the transfer conditions. As a result of the direct dry transfer technique, the graphene-polymer adhesion being stronger than graphene to $Si/SiO_2$ wafer.

## 1. Introduction

Graphene and other 2D materials have extraordinary characteristics that make them attractive for an uncountable number of applications, especially for flexible applications, i.e. applications where they will be integrated with polymers[1,2]. Advances have already happened on the direct synthesis of these materials and further improvements should still be expected but, since the low thermal stability of polymers, processes for the transfer of graphene onto polymeric surfaces will still be required. A process to transfer a graphene film from its original growth substrate to a given target surface is typically composed of different steps that include the coating of graphene with one or more support layers/films, the release of the graphene from its original substrate, the application of the graphene to the target substrate and the removal of the support films[3–5]. Different transfer processes have

Corresponding authors:
Tel.+55 11 21148077. Email address: guilherminojmf@mackenzie.br (G.J.M. Fechine)
Tel.+65 6516 6979. Email address: phyob@nus.edu.sg (B. Özyilmaz)

addressed the scalability[6–11], minimizing the defects or the residues from the support[7,9] or the residues from the etching/delamination[11] but the transfer process of graphene is still under discussion. In the particular case of the direct transfer of graphene to a target substrate, ref [12,13] demonstrated that, upon the use of an adhesive layer graphene may be mechanically peeled from the growth surface and, more recently, in ref [14] the authors showed that graphene could be put in contact with certain polymeric substrates though they would later chemically etch the catalyst metal to avoid graphene release from the polymer foils. So, the direct transfer of graphene to a polymer and the understanding of the mechanism governing this phenomenon have not been disclosed still. Besides, at this time, the expert's validation of a transfer is based on the full coverage of the target substrate with graphene for a later patterning process[15], possibly because most of the current applications are electronics oriented. But the development of applications targeting other functionalities of the graphene would benefit from a simpler process that would allow controlling graphene coverage at the transfer stage so that, for example, the overall degree of cellular differentiation in graphene deposited in different substrates[16], modification of surface activity of the catalysts by graphene[17] or optical response of graphene[18] of the (partially-) graphene-covered surface could be tuned. Also, defining full-non-transfer conditions (0% graphene transferred) should be of interest for molding/unmolding purposes. In this sense, it seems that the integration of graphene processing with conventional polymeric methods such as hotpressing, lamination, stamping or molding has still not been conceived.

Here, we study theoretically and experimentally the chemical interactions of the graphene-polymer systems and the rheological properties for different polymer cases and determine the keys for the direct transfer of graphene from its original growth substrate to a polymeric surface. We show that graphene transfer can be completed in a simple two-step process (polymer application and metal foil peeling) and that graphene coverage can be tuned from no-coverage (0% transfer) to full-coverage (100% transfer) by the only application of optimized heat and pressure conditions.

## 2. Experimental

### 2.1 Transfer of graphene

**Figure 1** shows the schematic of the transfer procedure. The method involves no materials but the graphene/metal foil and the polymer film, and the system to apply pressure and heat, in our case, a hotpress machine. In the first step (polymer application) the polymer is put in contact with the graphene and moderate pressure and temperature above the melting temperature of the polymer are applied in accordance with the transfer conditions. Once the stack is cooled down the metal foil is mechanically removed from the polymer/graphene (metal foil peeling). No chemical etching or electrochemical delamination process is needed.

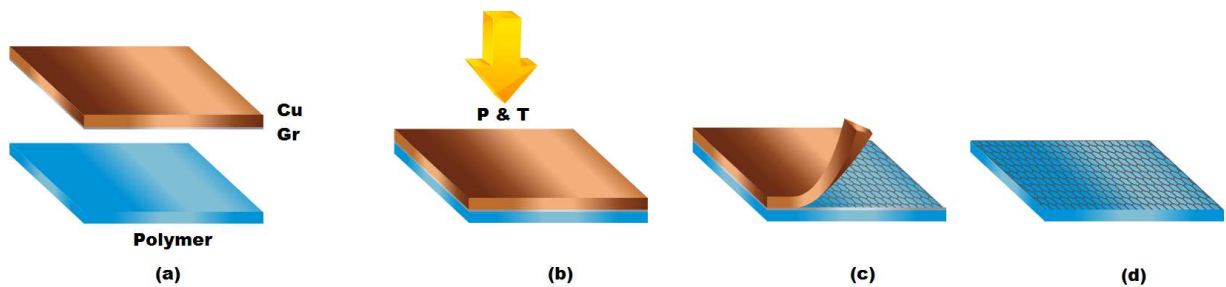

Figure 1 – Schematic of the transfer method and sample after transfer. (a-d) Schematic: (a) graphene/metal and polymer film before transfer. (b) Polymer application step to form the metal/graphene/polymer stack. (c) Peeling of the metal step. (d) Final graphene/polymer stack.

### 2.2 Materials

Low density and high density polyethylene (LDPE and HDPE), polystyrene (PS), poly(lactide acid) (PLA) and poly(vinylidenefluoride-co-trifluoroethylene, (70/30 PVDF-TrFE) were used for the transfers. This set of polymers was chosen because it represents different possibilities of graphene-polymer application, as flexible and inert (LDPE and HDPE), rigid and transparent (PS), biomedical

(PLA) and electronics (PVDF-TrFE) and because they have distinct values of surface energy. The preparation of the initial polymer foils is described in the Methods section. Single layer graphene grown by Chemical Vapor Deposition (CVD) on copper foil was always used as the graphene/metal source.

**2.3 Polymeric films preparation**

The polymer films were obtained by using the Hot Press Machine (HPM). Pellets of the polymers were used to produce the films. The pellets were placed between the plates of HPM until reaching the melting temperature. Films of 50 to 100 µm of thickness were obtained. These films were characterized by Differential Scanning Calorimetry - DSC (**Suppl. Inf. SI1)** and Oscillatory Rheological measurements.

**2.4 Direct dry transfer of graphene to polymeric substrates**

CVD-grown graphene is submitted to 120 ºC for 10 minutes and a nitrogen flow to eliminate water molecules from the surface. The polymeric films were washed with IPA (isopropyl alcohol) and submitted to nitrogen flow to clean the surface. CVD-grown graphene on copper foil and polymer film are placed between the two plates of the HPM. HPM is equipped with a heating/cooling system and pressure controller. The transfer is performed directly to the target substrate by melting the polymer and the application of moderate pressure. Melted polymer stays in contact with the graphene-copper foil for 5 minutes, after that the cooling system is turned on and graphene is transferred to the target substrate. At the end of this step, a set composed of polymer/graphene/copper is removed from the HPM. The copper foil is peeled out and polymer/graphene is obtained. The first choice of transfer temperature was set up ~10 to 20ºC higher than the melting temperature of the polymer to guarantee complete fusion. The temperature used for polystyrene (amorphous polymer) was estimated according to temperature of flow. Preliminary tests were completed to identify the first pressure where some graphene had been transferred to the polymer.

**2.5 Optical Images**

A 100x objective lens with Numerical aperture (NA) of 0.95 was used for the optical images. The improvement of contrast between polymer and graphene was only obtained with closing of the pin hole. A computational routine was built to analyze images of each graphene-polymer samples and quantity the area of the polymer covered by graphene (bright area).

**2.6 Raman Spectroscopy**

The Raman spectra were carried out with a WITEC CRM200 Raman system. The excitation source is a 534 nm laser with a laser power below 10 mW on the sample to avoid laser induced local heating. For the Raman image, the sample was placed on an x-y piezostage and scanned under the illumination of the laser. The Raman spectra from every part of the samples were recorded. Data analysis was completed using WITec Project software. The Raman Mapping area size was limited by the roughness of the graphene-polymer films since the focus is lost easily at long distances.

**2.7 Simulations**

Single layer graphene model was obtained from graphite crystal structure[19]. Random coil models of PE, PLA, PS and PVDF-TrFE were constructed, each comprising of 50 repeat units per chain. Low and high density polyethylene systems were represented as linear chains (PE). For PE, PLA, PS and PVDF-TrFE, 33, 11, 20 and 30 chains were randomly packed into a 3D cell respectively to generate a homogeneous amorphous polymer layer with an approximate thickness of 58 Å, at initial density of 0.6 g/cm$^3$ after which it was heated up (as described below) to achieve the equilibrium density. The equilibrium densities of PLA (1.20 g/cc), PS (0.97 g/cc) and PVDF-TrFE (1.80 g/cc) at room temperature are in good agreement with experiment while the density predicted for PE (0.77 g/cc) is representative of the low density polyethylene[20,21]. A single layer graphene was introduced in the 3D cell forming a polymer/graphene interface (see Supplementary Material – S5) and the Periodic Boundary Conditions (PBC) were applied to the graphene-polymer systems.

All simulations were carried out using the LAMMPS MD code[22], with the PCFF force field chosen to calculate interatomic interaction potentials[23]. PCFF, a polymer *ab initio* consistent force field that covers a broad range of organic materials has been shown to reproduce the density, free energy of binding and glass transition temperatures for a range of polymers, particularly those employed in this study[24–26]. All Molecular Dynamics simulations were carried out in the NVT (constant number of particles, volume and temperature) ensemble, with the PBC unit cell size of approximately 44 x 38 x 170 Å, extended in the Z direction to mimic an infinite polymer-graphene interface and to allow for thermal expansion of the systems at elevated temperature. The non-bonded interactions were evaluated with a cut off distance of 15.50 Å for the real space portion, with the particle-particle-particle-mesh solver (PPPM) method for the long-range electrostatics[27] and tail correction for the long range Van der Waals interactions. The Nose-Hoover thermostat was used to maintain a constant temperature[28]. The simulated graphene transfer comprised of three-steps (1-3). (1) First, the polymers were heated up to 211°C (which is above the melting temperature of all the polymers studied), ensuring the polymer density reached a steady state. This was typically achieved within the first nanosecond for all systems. (2) Second, graphene was introduced at a distance of 30 Å from the top surface of the melted polymer while applying a constant pressure of 350kPa in the negative z direction. The pressure was maintained for 4 ns, during which the position of the graphene layer was monitored. (3) Lastly, graphene - polymer system was cooled down to 25 °C at a rate of 50 °C/ns. To examine effects of the increased temperature, we repeated these steps (compression) with the temperature maintained at 25 °C.

**2.8 Oscillatory Rheology Tests**

The temperature sweep tests (2 °C min$^{-1}$) were performed within the linear viscoelastic regime under oscillatory shear at a frequency of 1 Hz and strain of 0.5% using a Hybrid-3 TA Instruments rheometer (plate-plate, diameter of 25 mm, gap of 500 µm).

## 3. Results and Discussion

Simulations using all-atom molecular dynamics were completed to mimic the experimental process and obtain an atomic-level understanding of physico-chemical factors affecting the transfer. Following the simulated graphene transfer process, we examined the adhesion strength of the formed graphene/polymer interface, by undertaking steered molecular dynamics simulations, retracting graphene from the polymer layer at a speed of 5 Å/ns with a force constant of 500 pN/Å coupling the two phases. During the retraction, the force was monitored as a function of vertical distance between the polymer and graphene (measured from the centre of mass of the polymer layer), from which the potential of mean force (PMF) was extracted. From these simulations, we ranked the relative strength of adhesion between the graphene layer and different polymers by estimating the force/ required to separate a single layer of graphene from the polymer surface at 25 and 211°C. To ensure the pristine (no polymer residues remaining) detachment of graphene from the film, we tethered the position of the polymer constituents, slightly restraining their movement in the z direction using a very soft force constant of 0.07 pN/Å. All simulations were performed with 1.0 fs time step and trajectories were generated by saving atomic coordinates in 10 ps time intervals. The duration of simulation for each polymer-graphene system spanned a total time of ~ 33 ns.

The calculated force curves between graphene and the tested polymers during retraction of the graphene layer are shown in **Figure 2**. Properties of the polymer-graphene systems examined, including the free energy of separation (FES), pull-off force (PF) and adhesion hysteresis (ΔE) between the high and room temperature treated systems are also shown in **Figure 2**. Simulations demonstrate that adhesion is hysteretic and depends not only on the history of the polymer treatment (e.g. high temperature *versus* room temperature) but also on the chemical composition of the polymer film. Three distinct consequences of heat treatment may be seen for all polymers: 1) the pull-off force (PF) is higher in magnitude as a result of heating, 2) the free energy of separation (FES), determined by the area underneath the force curve is also higher as a result of increased temperature and 3) with the exception of PE, the force curve is shifted towards shorter separations at high temperature, indicating a better contact between graphene and the polymer layer. The difference between the curves demonstrates that increasing the temperature during the transfer process results in a stronger force required to separate graphene from the polymer, which is manifested in adhesion hysteresis. PVDF-TrFE shows the highest values of the free energy of separation, pull-off force and adhesion hysteresis when compared with the other polymers tested. In comparison, PE and PLA demonstrate reduced adhesion with graphene as indicated by FES values. While these two polymers show similar adhesion strength with graphene, they exhibit remarkably different behaviors during retraction. PLA displays low flexibility driven by strong inter- and intramolecular hydrogen bonding. This results in a somewhat limited polymer elongation at detachment, as indicated by the relatively narrow width of the force curve, but a relatively high pull-off force (PF). In contrast, in PE the chains pack less closely. This is driven by relatively weaker dispersion forces and is manifested in a relatively low-detachment force (PF) and significant elongation upon detachment, with the force tending to zero over a greater distance compared to PLA. It is interesting to note that the simulations indicate that PS does not show a significant increase in adhesion to graphene following the treatment. This is likely due to the persistent internal π-π interactions between the aromatic rings, which render the polymer stiff and inflexible. Aromatic stacking can also contribute to a relatively strong interfacial interaction between polymer and graphene provided the polymer can reach the graphene surface in a favorable geometry. However, the graphene - PS interaction only slightly improves following the heat treatment thus confirming the effect of PS stiffness. Indeed, the results suggest that at elevated temperature the internal interactions still dominate in PS as demonstrated by the relatively low adhesion hysteresis and by the fact that the force curve exhibits only a slight shift towards the left upon treatment, indicating somewhat limited interfacial rearrangement during heating. These results indicate that adhesion between graphene and a pre-heated polymer is strongly influenced by the rheological properties of the polymer and thus the quality of graphene transfer is expected to depend not only on the processing conditions (temperature, pressure), but also on the viscoelastic properties of the polymer melt.

Good adhesion between graphene and polymer has to be higher to promote an excellent transfer of graphene to polymeric substrate. The adhesion between two substrates is a very complex phenomenon since it involves multidisciplinary knowledge, such physical chemistry of surfaces, chemical structure of the substrates and strength of materials[29]. Therefore, in the specific case here, the adhesion between the substrates is related with rheological properties of the polymer since the transfer of graphene starts above the melting temperature of the polymers.

Based on the suggested effects of the polymer rheology, a rheological study of the polymers has been completed to understand how the graphene transfer could be improved. Oscillatory rheology tests were performed at shear rate of 1 Hz and strain of 0.5% to simulate the low transfer process stress solicitation. A criterion for a material with good adhesion is a Storage Modulus (G´) value less than $10^5$ Pa, called Dahlquist criterion[30]. Materials exceeding the Dahlquist criterion have poor adhesion characteristics due to an inability to dissipate energy via viscous contributions, which means it cannot deform to make good contact with the substrate. Store Modules of polymers for all ranges of temperatures used for the transfers are in agreement with Dahlquist criterion (**Figure 3**).

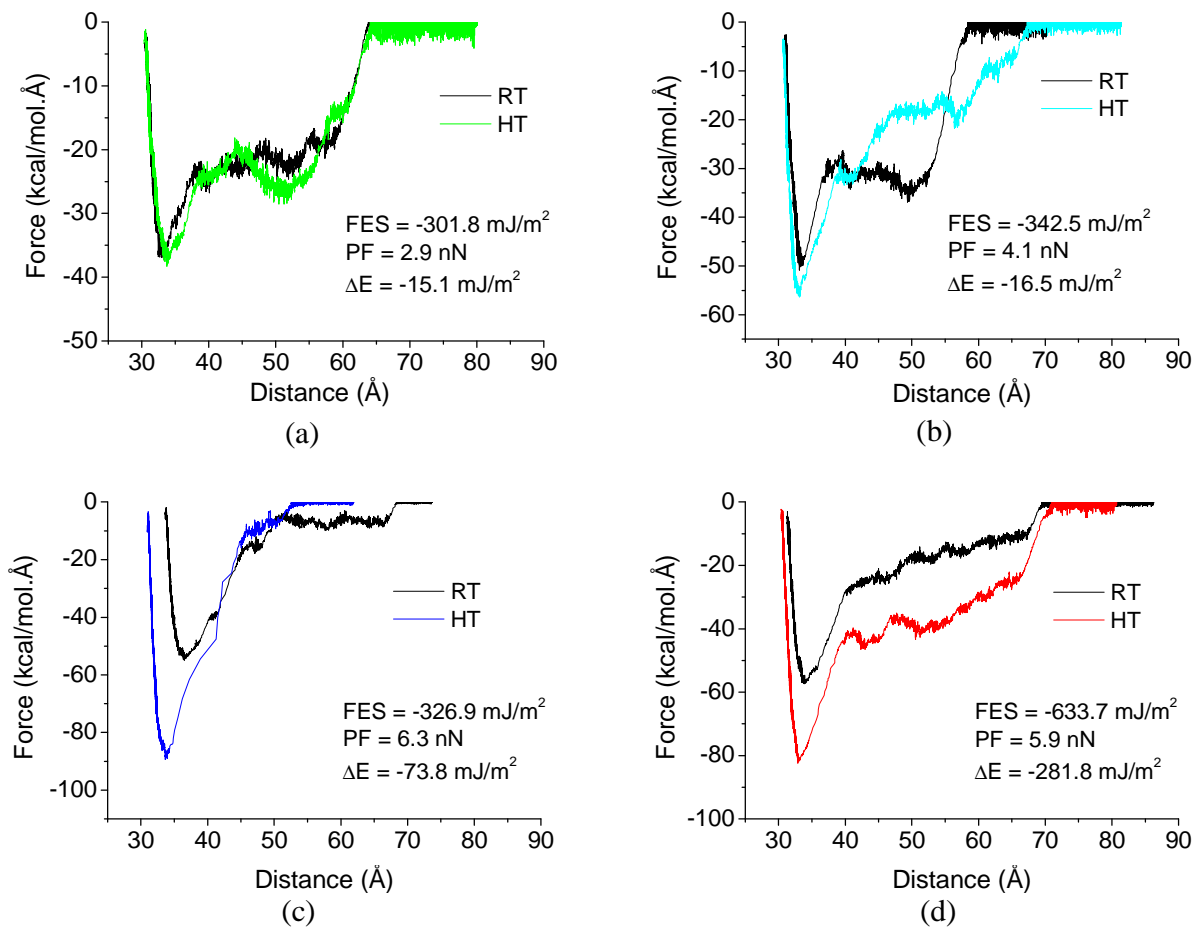

Figure 2 - Adhesion force between graphene and the different polymers and Free Energy of Separation (FES), Pull-off Force (PF) and Adhesion Hysteresis (ΔE) during retraction for untreated and thermally treated films. (a-d) PE, PS, PLA and PVDF-TrFE, respectively. RT: Room temperature, HT: High temperature.

Nevertheless, not all polymers that meet the Dahlquist criterion presented a good behavior for the graphene transfer at Condition 1 (**Figure 4**). PLA has a low G´ at 170 °C, but negligible Loss Modulus (G´´). PLA shows no dissipative modes even with low G´ and undergoes brittle failure; consequently, a poor adhesion is achieved. Another and no less important rheological property is the complex viscosity (η*). Solid substrates usually show degree of roughness which may reduce the contact with the polymer even if pressure is applied. The complex viscosity has to be low enough to allow the melted polymer to reach the valley of a substrate surface; therefore, sufficiently high to promote energy dissipation. PS has a low value of η* at 190 °C although it is not enough to keep a good adhesion with the graphene during the transfer.

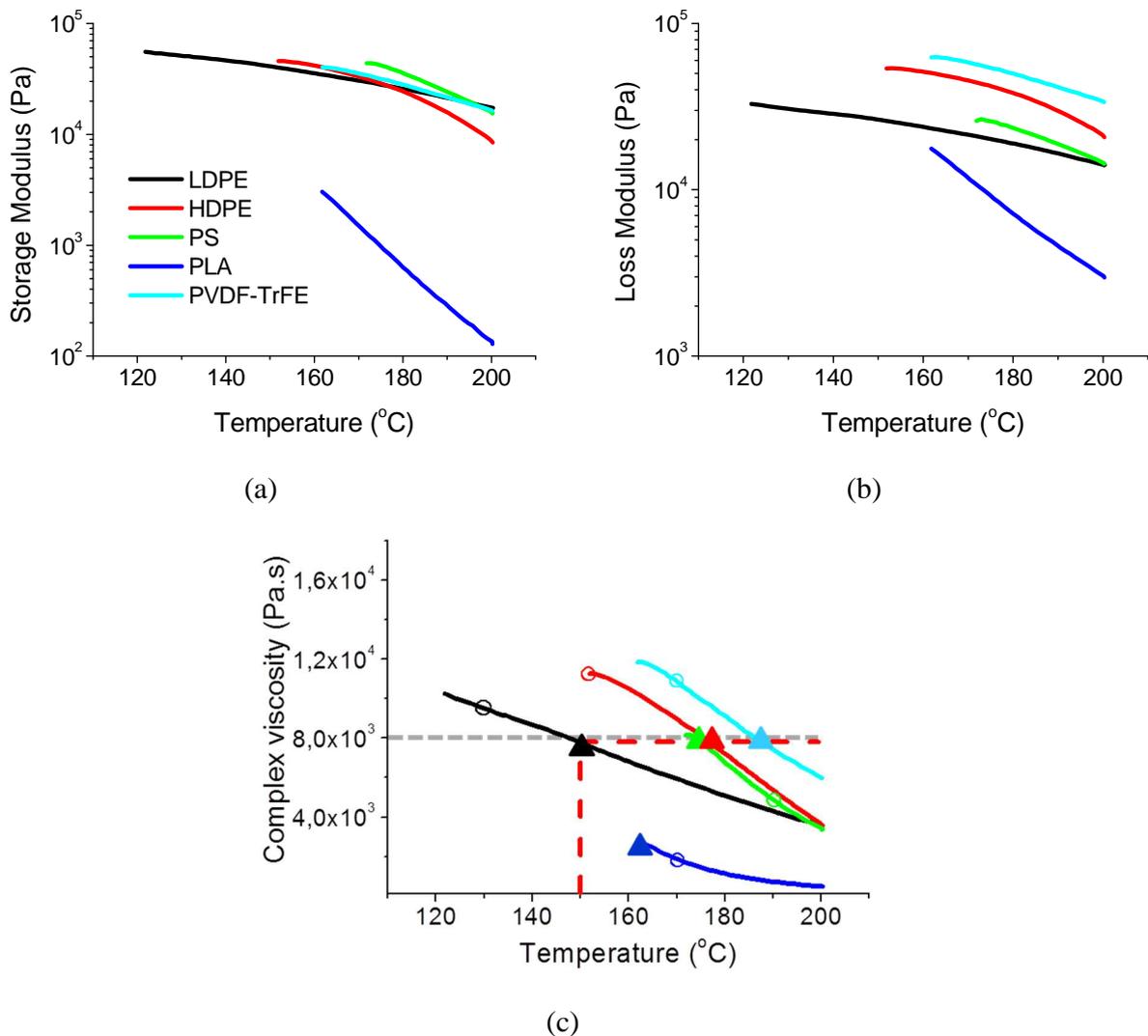

(a)

(b)

(c)

Figure 3 - Rheological characteristics of the graphene-polymer systems as function of temperature: (a) Storage Modulus; (b) Loss Modulus, and (c) Complex Viscosity. The circles and squares in (c) indicate transfer conditions in the first and second sets of transfers.

The graphene-LDPE system was chosen to evaluate the influence of the rheological properties and the process parameters on the transfer since it was the polymer that had the largest area covered by graphene at the Transfer Conditions 1 (**Figure 4**). The presence of graphene on the polymer after the transfer was confirmed by optical microscopy (bright area) and Raman spectroscopy. These samples showed that the amount of graphene transferred to LDPE increased when either the processing pressures and temperatures were increased or the roughness of the graphene/metal was decreased, the temperature showing the highest impact in the transfer yield (**Figure 4** and **Suppl Inf SI3**). After optimized temperature conditions (i.e. optimized graphene-polymer contact) full graphene transfer (100% graphene coverage) was achieved. These results demonstrate that the graphene-LDPE adhesion, $A_{p-g}$, was stronger than the binding of graphene to the growth substrate, $B_{g-Cu}$ (720 mJ/m$^2$ [13]). As a result of the graphene-LDPE adhesion being stronger than that of graphene to Si/SiO$_2$ wafer ($A_{g-SiW}$ ranges from 151 mJ/m$^2$ [31] to 450 mJ/m$^2$ [32]), graphene devices on our polymeric substrate would show a higher mechanical stability compared to the other case. After these results, the η* of LDPE at 150°C (8·10$^3$ Pa·s) was chosen as the optimal rheological condition for an improved graphene-polymer contact for this particular graphene/metal case.

To validate the previous results we completed two sets of transfers at two distinctive conditions with the other graphene-polymer systems: HDPE, PS, PLA and PVDF-TrFE. In the first case, the temperature was set 10-20°C above the melting temperature of the polymer. In the second case, the temperature was set to match the rheological conditions for an improved graphene-polymer contact (η* = 8·10$^3$ Pa·s). Since η* of PLA could not be made to match the optimized value, in this case the transfer temperature was set to the melting temperature of the polymer. Figure 4 compares the optical images and compiles the Raman spectra and maps of the 2D peaks of graphene from the graphene-polymer substrates after the transfers. Again, the graphene appears like brighter areas when imaged under the optical microscope. The first observation is that the outcome of the transfer is strongly sensitive to the processing conditions. When tuning the temperature for an optimized graphene-polymer contact (optimized η* conditions) almost full-graphene-transfer, 95%, 80% and 99%, was again achieved in the HDPE, PLA and PVDF-TrFE cases, respectively. It is also remarkable that there was no graphene transferred (0% graphene coverage) in the PS case after the first transfer conditions. Therefore, the adhesion force between PS and graphene needs to be well below that of graphene to the substrate. This result is very positive since it means that graphene could have a big impact, for example, in the unmolding step in molding like processing. Last, the Raman analysis of the graphene shows that the mono and bi-layer graphene was transferred without introducing damage as the D/G ratios are low, is indicative of good quality single layer graphene[33] (Figures 4 and **Table 1**). In conclusion, these results show that the efficiency of the transfer of graphene transfer can be tuned by controlling the viscosity of the melted polymer.

Table 1 – Raman characteristics of the graphene after transfer to the different polymers.

| Polymer | G band [cm$^{-1}$] | 2D band [cm$^{-1}$] | FWHM (2D) [cm$^{-1}$] | 2D/G | D/G |
|---|---|---|---|---|---|
| LDPE | 1585.4 | 2676.6 | 36.5 | 3.8 | 0.1 |
| HDPE | 1585.2 | 2682.7 | 34.3 | 4.7 | 0.0 |
| PS | 1584.1 | 2686.3 | 32.5 | 1.4 | 0.2 |
| PLA | 1591.1 | 2681.8 | 31.1 | 2.3 | 0.1 |
| PVDF-TrFE | 1592.1 | 2682.8 | 35.4 | 1.2 | 0.2 |

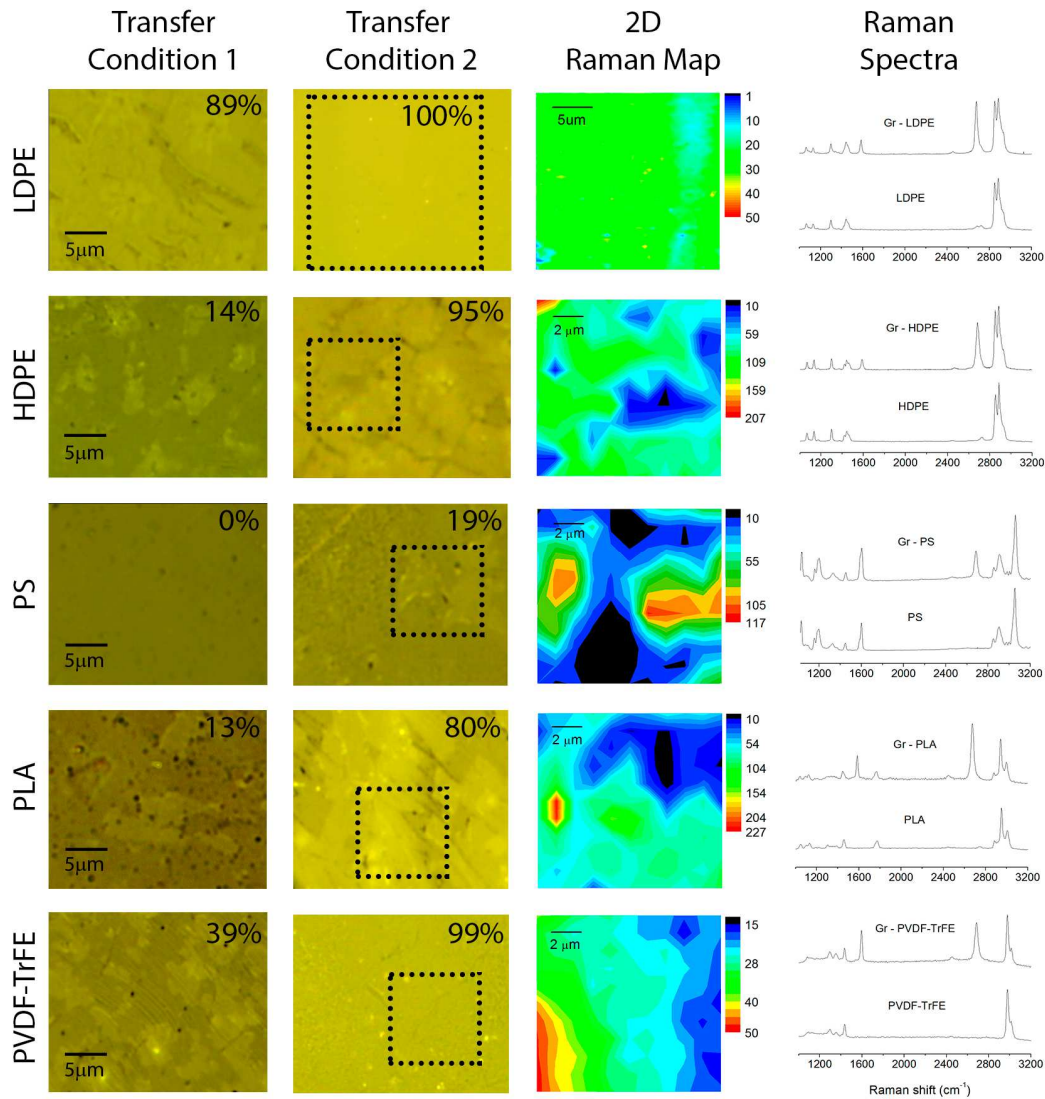

Figure 4 – Characterization of the graphene/polymer systems after transfer at different conditions. Each row refers to one polymer. (column 1 and 2) Optical pictures after transfer conditions 1 and 2, respectively. The amount of transferred graphene in each case is highlighted in the top right corner of the optical images. (column 3) Raman maps of the 2D band of graphene of the 625 µm$^2$ for LDPE and 100 µm$^2$ for the others polymers - areas highlighted in column 2. (column 4) Representative Raman spectra of the graphene on the different polymers.

## 4. Summary

Our theoretical and experimental work demonstrates that the proper understanding of the graphene-polymers interface together with the rheological characteristics of the graphene-polymer system are the key to a direct transfer of graphene to a given polymer. Our experiments show that the graphene transfer degree (the graphene covered area) can be controlled by the only application of distinctive temperature and pressure conditions. We have also shown that, after optimized graphene-polymer contact conditions, a full-transfer or non-transfer of graphene to the polymer (100% or 0% coverage, respectively) is possible. Our capability to control the graphene transfer degree is fundamental since different applications may require different graphene coverage. For example, electronic-like applications should require graphene continuity, thus full coverage, whereas bio-oriented applications could require partial coverage, and molding processes could benefit from no-transfer conditions. As a result of the direct dry transfer technique the graphene-polymer adhesion being stronger than graphene to Si/SiO$_2$ wafer (> 450 mJ/m$^2$). These results are of interest to any polymer-graphene system. Finally, this methodology being applicable to any polymeric liquid and to any polymeric method beyond a hotpress, our results are relevant to any transfer method and/or application involving the interface of a 2D material with a polymer.

## Acknowledgements

GJMF acknowledge the financial support from Brazilian funding agency (FAPESP 2012/50259-8 and 2010/17197-9).

GY and IY acknowledge the generous allocation of high performance computational resources from the Australian National Computational Infrastructure (NCI), the Western Australian computational facility (iVEC) and the Victorian Partnership for Advanced Computing (VPAC).

# References

[1]   Novoselov KS, Fal'ko VI, Colombo L, Gellert PR, Schwab MG, Kim K. A roadmap for graphene. Nature 2012;490:192–200. doi:10.1038/nature11458.

[2]   Butler SZ, Hollen SM, Cao L, Cui Y, Gupta JA, Gutie HR, et al. Progress, Challenges, and Opportunities in Two-Dimensional Materials Beyond Graphene. ACS Nano 2013;7:2898–926.

[3]   Kang J, Shin D, Bae S, Hong BH. Graphene transfer: key for applications. Nanoscale 2012;4:5527–37. doi:10.1039/c2nr31317k.

[4]   Chen X-D, Liu Z-B, Zheng C-Y, Xing F, Yan X-Q, Chen Y, et al. High-quality and efficient transfer of large-area graphene films onto different substrates. Carbon N Y 2013;56:271–8. doi:10.1016/j.carbon.2013.01.011.

[5]   Suk JW, Kitt A, Magnuson CW, Hao Y, Ahmed S, An J, et al. Transfer of CVD-grown monolayer graphene onto arbitrary substrates. ACS Nano 2011;5:6916–24. doi:10.1021/nn201207c.

[6]   Unarunotai S, Koepke JC, Tsai C-L, Du F, Chialvo CE, Murata Y, et al. Layer-by-layer transfer of multiple, large area sheets of graphene grown in multilayer stacks on a single SiC wafer. ACS Nano 2010;4:5591–8. doi:10.1021/nn101896a.

[7]   Verma VP, Das S, Lahiri I, Choi W. Large-area graphene on polymer film for flexible and transparent anode in field emission device. Appl Phys Lett 2010;96:203108. doi:10.1063/1.3431630.

[8]   Bae S, Kim H, Lee Y, Xu X, Park J-S, Zheng Y, et al. Roll-to-roll production of 30-inch graphene films for transparent electrodes. Nat Nanotechnol 2010;5:574–8. doi:10.1038/nnano.2010.132.

[9]   Wang D-Y, Huang I-S, Ho P-H, Li S-S, Yeh Y-C, Wang D-W, et al. Clean-lifting transfer of large-area residual-free graphene films. Adv Mater 2013;25:4521–6. doi:10.1002/adma.201301152.

[10]  Gao L, Ni G-X, Liu Y, Liu B, Castro Neto AH, Loh KP. Face-to-face transfer of wafer-scale graphene films. Nature 2014;505:190–4. doi:10.1038/nature12763.

[11]  Cherian CT, Giustiniano F, Martin-Fernandez I, Andersen H, Balakrishnan J, Ozyilmaz B. "Bubble-free" electrochemical delamination of CVD graphene films. Small 2014. doi:10.1002/smll.201402024.

[12]  Lock EH, Baraket M, Laskoski M, Mulvaney SP, Lee WK, Sheehan PE, et al. High-quality uniform dry transfer of graphene to polymers. Nano Lett 2012;12:102–7. doi:10.1021/nl203058s.

[13]   Yoon T, Shin WC, Kim TY, Mun JH, Kim T-S, Cho BJ. Direct measurement of adhesion energy of monolayer graphene as-grown on copper and its application to renewable transfer process. Nano Lett 2012;12:1448–52. doi:10.1021/nl204123h.

[14]   Martins LGP, Song Y, Zeng T, Dresselhaus MS, Kong J, Araujo PT. Direct transfer of graphene onto flexible substrates. Proc Natl Acad Sci U S A 2013;110:17762–7. doi:10.1073/pnas.1306508110.

[15]   Song J, Kam F-Y, Png R-Q, Seah W-L, Zhuo J-M, Lim G-K, et al. A general method for transferring graphene onto soft surfaces. Nat Nanotechnol 2013;8:356–62. doi:10.1038/nnano.2013.63.

[16]   Akhavan O, Ghaderi E, Shahsavar M. Graphene nanogrids for selective and fast osteogenic differentiation of human mesenchymal stem cells. Carbon N Y 2013;59:200–11. doi:10.1016/j.carbon.2013.03.010.

[17]   Kou R, Shao Y, Wang D, Engelhard MH, Kwak JH, Wang J, et al. Enhanced activity and stability of Pt catalysts on functionalized graphene sheets for electrocatalytic oxygen reduction. Electrochem Commun 2009;11:954–7. doi:10.1016/j.elecom.2009.02.033.

[18]   Fang Z, Thongrattanasiri S, Schlather A, Liu Z, Ma L, Wang Y, et al. Gated tunability and hybridization of localized plasmons in nanostructured graphene. ACS Nano 2013;7:2388–95. doi:10.1021/nn3055835.

[19]   McKie C, McKie D. Essentials of Crystallography. Blackwell Scientific Publications; 1986.

[20]   Navid A, Lynch CS, Pilon L. Purified and Porous Poly(Vinylidene Fluoride-Trifluoroethylene) Thin Films For Pyroelectric Infrared Sensing and Energy Harvesting. Smart Mater Struct 2010;19:055006.

[21]   Brandrup J, Immergut EH, Grulke EA, Abe A, Bloch DR. Polymer handbook. 1999.

[22]   Plimpton S. Fast parallel algorithms for short-range molecular dynamics.pdf. J Comput Phys 1995;117:1–19.

[23]   Sun H, Mumby SJ, Maple JR, Hagler AT. An ab Initio CFF93 All-Atom Force Field for Polycarbonates. J Am Chem Soc 1994;116:2978–87.

[24]   Karst D, Yang Y. Molecular modeling study of the resistance of PLA to hydrolysis based on the blending of PLLA and PDLA. Polymer (Guildf) 2006;47:4845–50. doi:10.1016/j.polymer.2006.05.002.

[25]   Shenogin S, Ozisik R. Deformation of glassy polycarbonate and polystyrene: the influence of chemical structure and local environment. Polymer (Guildf) 2005;46:4397–404. doi:10.1016/j.polymer.2005.03.015.

[26]   Hu M, Keblinski P, Li B. Thermal rectification at silicon-amorphous polyethylene interface. Appl Phys Lett 2008;92:211908. doi:10.1063/1.2937834.

[27]   Hockney RW, Eastwood JW. Computer Simulation Using Particles. Taylor & Francis; 1989.

[28]   Nosé S. A unified formulation of the constant temperature molecular dynamics methods. J Chem Phys 1984;81:511. doi:10.1063/1.447334.

[29]   Fourche G. An overview of the basic aspects of polymer adhesion. Part I: Fundamentals. Polym Eng Sci 1995;35:957–67. doi:10.1002/pen.760351202.

[30]   Zosel a. The effect of bond formation on the tack of polymers. J Adhes Sci Technol 1997;11:1447–57. doi:10.1163/156856197X00237.

[31]   Zong Z, Chen C-L, Dokmeci MR, Wan K. Direct measurement of graphene adhesion on silicon surface by intercalation of nanoparticles. J Appl Phys 2010;107:026104. doi:10.1063/1.3294960.

[32]   Koenig SP, Boddeti NG, Dunn ML, Bunch JS. Ultrastrong adhesion of graphene membranes. Nat Nanotechnol 2011;6:543–6. doi:10.1038/nnano.2011.123.

[33]   Ferrari AC, Meyer JC, Scardaci V, Casiraghi C, Lazzeri M, Mauri F, et al. Raman Spectrum of Graphene and Graphene Layers. Phys Rev Lett 2006;97:187401.

# SUPPLEMENTARY INFORMATION

**SI1 - Polymer film characterization**

Polymer films were characterized by Differential Scanning Calorimetry (DSC) before the graphene transfer to verify their thermal behavior (Figure SI1). DSC analyses were performed with a TA instrument 2920 equipment at a heating rate of 10ºC/min. Figure S4a, S4b and S4c show melting temperature to LDPE, HDPE and PLA, respectively. Melting temperature of polystyrene could not be seen on DSC curve (Figure SI1d) because it is an amorphous polymer. Only an inflection placed at 100ºC is noted due to glass temperature. DSC curve of PVDF-TrFE shows two thermal transitions (Figure SI1e): a) Curie Temperature and b) Melting temperature.

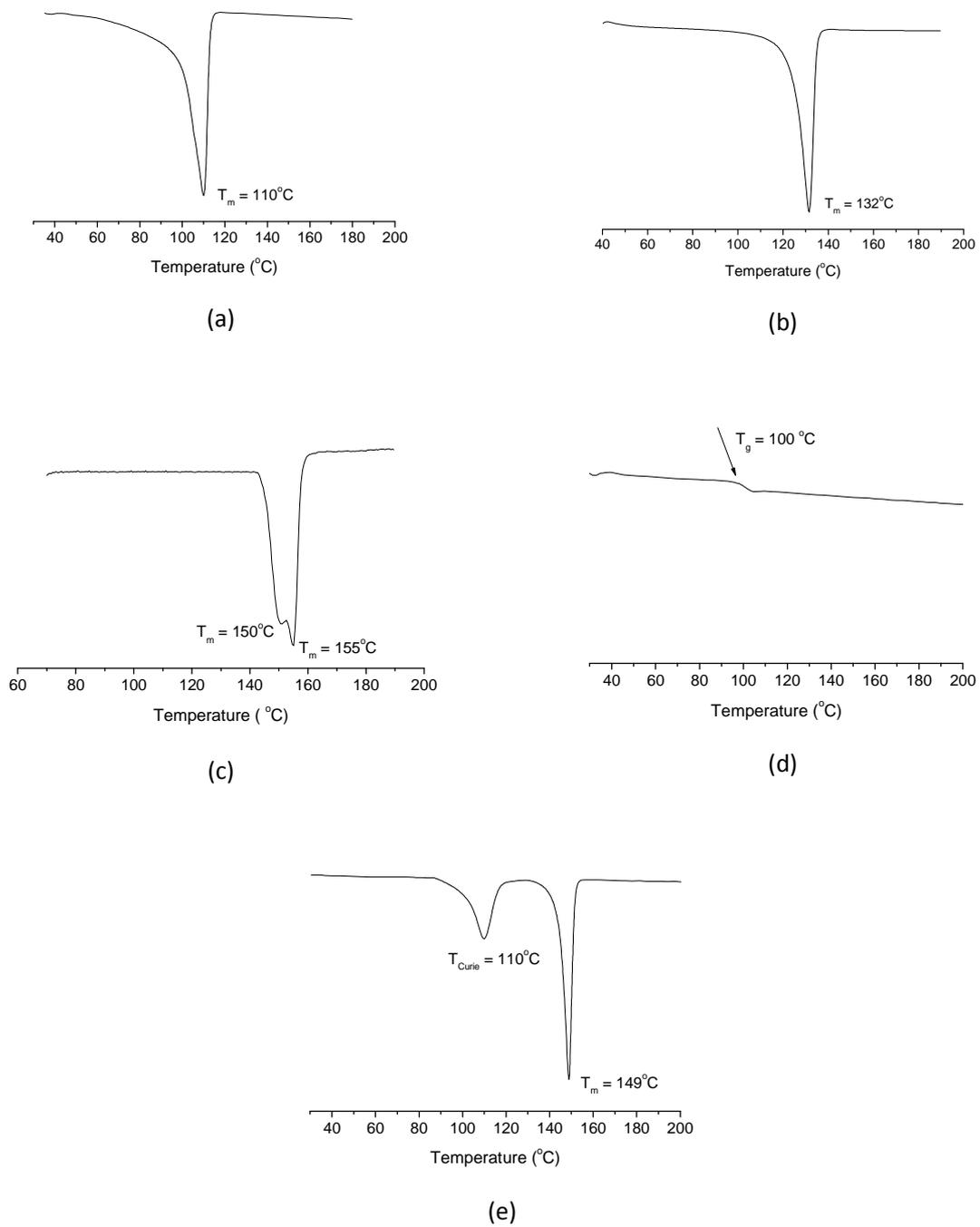

Figure SI1 – Thermal analysis curves (DSC) of LDPE (a), HDPE (b), PLA (c), PS (d) and PVDF-TrFE (e).

**SI2 - Simulation**

Figure SI2 show the schematic approach used on simulations as a model to represent a single layer of graphene and a random coil model PE comprising of 50 repeat units per chain polymer.

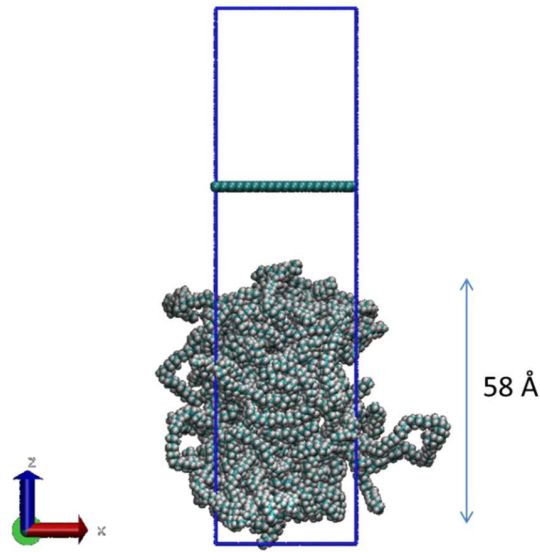

Figure SI2 - Schematic representation of polyethylene - graphene composite with the unit cell displaying dimensions of 44.28 Å in X, 38.35 Å in Y and 170 Å in Z.

**SI3 – Evaluation of experimental parameters on graphene transfer to LDPE**

Figure SI3 shows optical microscopy images and Raman mapping of graphene transferred to LDPE using two different sources of CVD graphene: high and low roughness, as well as two levels of temperature and pressure. Data presented in Figure SI3 indicate that the quantity of graphene transferred to LDPE increases with a decrease in roughness of CVD-graphene copper foils and an increase in temperature and pressure. Remarkably, the complete coverage of LDPE by graphene was achieved when using high levels of temperature and pressure with a low roughness CVD graphene copper foil (Figure SI3d).

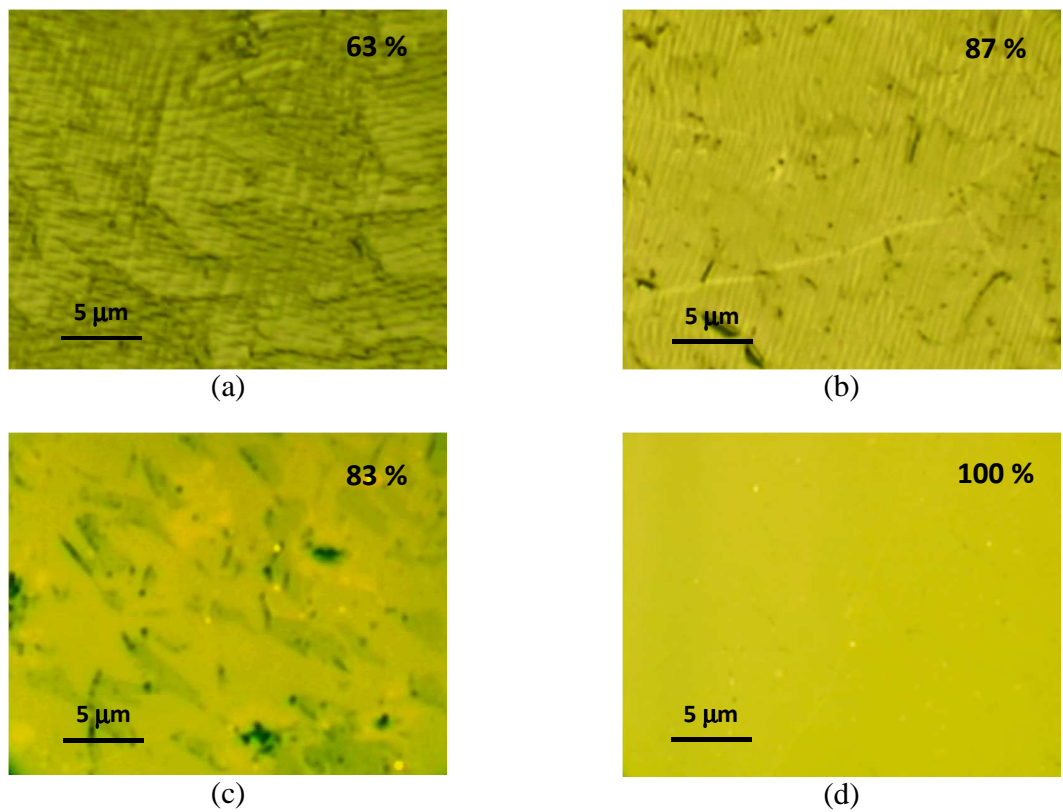

Figure SI3 - Optical images and Raman spectra of graphene transferred to LDPE under different conditions. Low Pressure, Low Temperature and High roughness of CVD graphene (a), High Pressure, Low Temperature and High roughness of CVD graphene (b), High Pressure, Low Temperature and Low roughness of CVD graphene (c), High Pressure, High Temperature and Low roughness of CVD graphene (d). Quantities of graphene transferred (%) to the polymers are shown in the upper right corner of the optical images.